\documentclass[aps,prb,superscriptaddress,twocolumn,10pt,longbibliography]{revtex4-2}
\usepackage{amsmath, amssymb}
\usepackage{braket}
\usepackage{color}
\usepackage{graphicx}
\usepackage{ulem}
\usepackage{siunitx}

\begin{document}
\date{\today}
\title{\textit{Ab initio} study of magnetoresistance effect in $\mathrm{Mn_{3}Sn}/\mathrm{MgO}/\mathrm{Mn_{3}Sn}$ antiferromagnetic tunnel junction}
\author{Katsuhiro Tanaka}
\affiliation{Department of Physics, University of Tokyo, Hongo, Bunkyo-ku, Tokyo 113-0033, Japan}
\author{Yuta Toga}
\affiliation{JSR-UTokyo Collaboration Hub CURIE, JSR Corporation, Higashi-Shinbashi, Minato-ku, Tokyo 105-8640, Japan}
\author{Susumu Minami}
\affiliation{Department of Physics, University of Tokyo, Hongo, Bunkyo-ku, Tokyo 113-0033, Japan}
\affiliation{Department of Mechanical Engineering and Science, Kyoto University, Kyoto 615-8540, Japan}
\author{Satoru Nakatsuji}
\affiliation{Department of Physics, University of Tokyo, Hongo, Bunkyo-ku, Tokyo 113-0033, Japan}
\affiliation{Trans-Scale Quantum Science Institute, University of Tokyo, Hongo, Bunkyo-ku, Tokyo 113-0033, Japan}
\affiliation{Institute for Solid State Physics, University of Tokyo, Kashiwa, Chiba 277-8581, Japan}
\affiliation{Institute for Quantum Matter, Department of Physics and Astronomy, Johns Hopkins University, Baltimore, Maryland 21218, USA}
\author{Takuya Nomoto}
\affiliation{Department of Physics, Tokyo Metropolitan University, Hachioji, Tokyo 192-0397, Japan}
\author{Takashi Koretsune}
\affiliation{Department of Physics, Tohoku University, Sendai, Miyagi 980-8578, Japan}
\author{Ryotaro Arita}
\affiliation{Department of Physics, University of Tokyo, Hongo, Bunkyo-ku, Tokyo 113-0033, Japan}
\affiliation{Trans-Scale Quantum Science Institute, University of Tokyo, Hongo, Bunkyo-ku, Tokyo 113-0033, Japan}
\affiliation{Center for Emergent Matter Science, RIKEN, Wako, Saitama 351-0198, Japan}
\begin{abstract}
The antiferromagnets with the time-reversal symmetry broken magnetic structures possess a finite spin splitting in the momentum space,
and may contribute to a realization of a finite tunnel magnetoresistance (TMR) effect even with magnets with zero net spin polarization.
In this paper, we study the TMR effect with the noncollinear antiferromagnet $\mathrm{Mn_{3}Sn}$ whose inverse $120^{\circ}$ antiferromagnetic order breaks the time-reversal symmetry.
In particular, we employ the representative barrier material $\mathrm{MgO}$ as the tunnel insulator, and calculate the TMR effect in the $\mathrm{Mn_{3}Sn}(01\bar{1}0)/\mathrm{MgO}(110)/\mathrm{Mn_{3}Sn}$ magnetic tunnel junctions (MTJs),
which has an optimal geometry for the spin-orbit torque switching of the magnetic configurations.
We show that a finite TMR ratio reaching $\gtrsim 1000${\%} appears in the $\mathrm{Mn_{3}Sn}/\mathrm{MgO}/\mathrm{Mn_{3}Sn}$ MTJs,
which is due to the spin splitting properties of $\mathrm{Mn_{3}Sn}$ in the momentum space combined with the screening effect of $\mathrm{MgO}$.
\end{abstract}
\maketitle
\section{Introduction}
\label{sec:introduction}
The magnetic tunnel junction (MTJ) is a multilayer system consisting of magnetic metal electrodes and an insulating barrier in between.
The MTJs have been used in a nonvolatile memory storage, magnetic random access memory (MRAM).
Basically, the MTJ can take two distinct magnetic states, which correspond to `0' and `1' bit states;
the relative directions of the magnetic moments of two electrodes in the MTJs are parallel or antiparallel.
\par
To read the information stored in MTJs, the tunnel magnetoresistance (TMR) effect~\cite{Julliere1975_PhysLettA_54A_225} is used.
The TMR effect is a spin-dependent transport phenomenon;
the tunneling current through the MTJ can be different depending on the relative directions of the magnetic moments of the two magnetic electrodes,
namely, the current can be different between the `0' and `1' states.
This difference is utilized to extract the information from MTJs.
The origin of the TMR effect is the spin-polarization of the tunneling electrons,
and thus the TMR effect has been typically discussed with ferromagnets,
where we can naturally have the tunneling electrons with a finite spin-polarization~\cite{Julliere1975_PhysLettA_54A_225,Miyazaki1995_JMagnMagnMater_139_L231,Moodera1995_PhysRevLett_74_3273,Butler2001_PhysRevB_63_054416,Mathon2001_PhysRevB_63_220403,Parkin2004_NatMater_3_862,Yuasa2004_NatMater_3_868,Scheike2023_ApplPhysLett_122_112404,Tsymbal2003_JPhysCondensMatter_15_R109,Zhang2003_JPhysCondensMatter_15_R1603,Ito2006_JMagnSocJpn_30_1,Yuasa2007_JPhysD_40_R337,Butler2008_SciTechnolAdvMater_9_014106}.
Here, the so-called Julliere picture~\cite{Julliere1975_PhysLettA_54A_225} has been used as a convenient approach to grasp the TMR effect;
the net spin polarization of the magnetic electrodes plays a role in generating a finite TMR effect, 
while the coherent tunneling mechanism has been proposed for a more microscopic understanding~\cite{Butler2001_PhysRevB_63_054416,Mathon1997_PhysRevB_56_11810,Butler2008_SciTechnolAdvMater_9_014106}.
\par
By contrast, the TMR effect using antiferromagnetic electrodes has been discussed recently~\cite{Merodio2014_ApplPhysLett_105_122403,Stamenova2017_PhysRevB_95_060403,Zelezny2017_PhysRevLett_119_187204,Jia2020_SciChinnaPhys_63_297512,Shao2021_NatCommun_12_7061,Smejkal2022_PhysRevX_12_011028,Dong2022_PhysRevLett_128_197201,Chen2023_Nature_613_490,Qin2023_Nature_613_485,Shao2023_PhysRevLett_130_216702,Tanaka2023_PhysRevB_107_214442,Cui2023_PhysRevB_108_024410,Jia2023_PhysRevB_108_104406,Jiang2023_PhysRevB_108_174439,Shi2024_AdvMater_36_2312008,Chi2024_PhysRevApplied_21_034038,Samanta2024_PhysRevB_109_174407,Zhu2024_JMagnMagnMater_597_172036,Zhu2024_ChinPhysLett_41_047502,Tanaka2024_PhysRevB_110_064433,Chou2024_NatCommun_15_7840,Gurung2024_NatCommun_15_10242,Wang2024_ApplPhysLett_125_202404,Luo2025_PhysRevB_111_144417,Yang2025_Newton_1_100142,Liu2025_AdvSci_12_e02985,Noh2025_PhysRevLett_134_246703,Zhu2025_ApplPhysLett_127_082401,Sun2025_PhysRevB_112_094411,Yang2025_arXiv_2505.17192,Kang2025_arXiv_2509.03026,Shao2024_npjSpintronics_2_13,Tanaka2025_JPhysCondensMatter_37_183003};
while the conventional Julliere's picture for antiferromagnets does not give a TMR effect due to the absence of net spin polarization,
the characteristic magnetic structures of antiferromagnets allow us to realize the TMR effect.
Particularly, in the antiferromagnets whose magnetic orders break the time-reversal symmetry like in noncollinear antiferromagnets~\cite{Nakatsuji2022_AnnuRevCondensMatterPhys_13_119,Rimmler2024_NatRevMater} or altermagnets~\cite{Smejkal2022_PhysRevX_12_031042,Smejkal2022_PhysRevX_12_040501},
a finite spin-splitting in the momentum space appears. 
Utilizing this spin polarization, antiferromagnetic metals can also generate an intrinsic spin-polarized tunneling current in the MTJs, 
which produces a finite TMR effect ~\cite{Zelezny2017_PhysRevLett_119_187204,Shao2021_NatCommun_12_7061,Smejkal2022_PhysRevX_12_011028,Dong2022_PhysRevLett_128_197201,Chen2023_Nature_613_490,Qin2023_Nature_613_485,Shao2023_PhysRevLett_130_216702,Cui2023_PhysRevB_108_024410,Jiang2023_PhysRevB_108_174439,Shi2024_AdvMater_36_2312008,Chi2024_PhysRevApplied_21_034038,Samanta2024_PhysRevB_109_174407,Zhu2024_ChinPhysLett_41_047502,Tanaka2024_PhysRevB_110_064433,Chou2024_NatCommun_15_7840,Gurung2024_NatCommun_15_10242,Wang2024_ApplPhysLett_125_202404,Luo2025_PhysRevB_111_144417,Yang2025_Newton_1_100142,Liu2025_AdvSci_12_e02985,Noh2025_PhysRevLett_134_246703,Zhu2025_ApplPhysLett_127_082401,Sun2025_PhysRevB_112_094411,Yang2025_arXiv_2505.17192,Kang2025_arXiv_2509.03026}.
The TMR effect with such antiferromagnets breaking the time-reversal symmetry has been observed in experiments using noncollinear antiferromagnets~\cite{Chen2023_Nature_613_490,Qin2023_Nature_613_485,Shi2024_AdvMater_36_2312008,Chou2024_NatCommun_15_7840,Kang2025_arXiv_2509.03026,Kang2025_arXiv_2509.03026} and altermagnets~\cite{Noh2025_PhysRevLett_134_246703} as well as the theoretical proposals.
\par
In this paper, we study the TMR effect with a kagome noncollinear antiferromagnet, $\mathrm{Mn_{3}Sn}$ \cite{Nagamiya1982_SolidStateCommun_42_385,Tomiyoshi1982_JPhysSocJpn_51_803,Brown1990_JPhysCondensMatter_2_9409,Nakatsuji2015_Nature_527_212}.
$\mathrm{Mn_{3}Sn}$ takes the inverse $120^{\circ}$ antiferromagnetic structure at room temperature, 
which breaks the time-reversal symmetry macroscopically.
This time-reversal symmetry broken magnetic structure has led to an experimental observation of the TMR effect~\cite{Chen2023_Nature_613_490,Chou2024_NatCommun_15_7840} as well as several physical properties typically discussed with ferromagnets~\cite{Kubler2014_EPL_108_67001,Nakatsuji2015_Nature_527_212,Ikhlas2017_NatPhys_13_1085,Higo2018_NatPhoton_12_73,Chen2021_NatCommun_12_572,Nakatsuji2022_AnnuRevCondensMatterPhys_13_119,Higo2022_JMagnMagnMater_564_170176} due to the topologically nontrivial electronic structure~\cite{Kuroda2017_NatMater_16_1090,Chen2021_NatCommun_12_572}.
This motivates us to utilize $\mathrm{Mn_{3}Sn}$ as a central part of the antiferromagnetic spintronic devices,
which will have advantages compared to the ferromagnetic spintronic devices such as the less stray field or the faster switching speed~\cite{Miwa2021_SmallSci_1_2000062,MacDonald2011_PhilTransRSocA_369_3098,Gomonay2014_LowTempPhys_40_17,Jungwirth2016_NatNanotechnol_11_231,Baltz2018_RevModPhys_90_015005,Zelezny2018_NatPhys_14_220,Siddiqui2020_JApplPhys_128_040904,Amin2020_ApplPhysLett_117_010501,Fukami2020_JApplPhys_128_070401,Rimmler2024_NatRevMater,Chen2024_AdvMater_36_2310379}.
\par
Particularly, here we focus on the TMR effect in the $\mathrm{Mn_{3}Sn}$-MTJ with the $\mathrm{Mn_{3}Sn}(01\bar{1}0)$ orientation.
With this geometry, the electrical switching of the magnetic moments of $\mathrm{Mn_{3}Sn}$ via the spin-orbit torque has been realized experimentally~\cite{Higo2022_Nature_607_474,Yoon2023_NatMater_22_1106,Takeuchi2025_Science_389_8301},
and the reduction of the electric current density required to switch the magnetic moments compared to other geometries is also suggested theoretically~\cite{Xu2024_PhysRevB_109_134433}.
Hence, aiming at the application to the electrical operations of the $\mathrm{Mn_{3}Sn}$-based spintronic devices,
this $\mathrm{Mn_{3}Sn}(01\bar{1}0)$ geometry is suitable.
With $\mathrm{MgO}$, a typical barrier material for MTJs,
we construct $\mathrm{Mn_{3}Sn}(01\bar{1}0)/\mathrm{MgO}(110)/\mathrm{Mn_{3}Sn}$ MTJs.
We perform the calculations of the tunneling conductance from first principles
and obtain a finite TMR effect, where the TMR ratio reaches $\gtrsim$ 1000\%.
By combining the analysis of the spin polarization of bulk $\mathrm{Mn_{3}Sn}$ and the complex band structure of $\mathrm{MgO}$ with the TMR calculations,
we discuss that the TMR effect in the $\mathrm{Mn_{3}Sn}/\mathrm{MgO}/\mathrm{Mn_{3}Sn}$ MTJs occurs reflecting the momentum-dependent spin splitting in $\mathrm{Mn_{3}Sn}$ and the screening of tunneling electrons by $\mathrm{MgO}$.
\par
Additionally, this study will have a practical importance for the application of $\mathrm{Mn_{3}Sn}$ to magnetic electrodes of MTJs.
Several studies have discussed the TMR effect using $\mathrm{Mn_{3}Sn}$ theoretically~\cite{Chen2023_Nature_613_490,Dong2022_PhysRevLett_128_197201,Wang2024_ApplPhysLett_125_202404}.
These previous studies, however, have considered the MTJs without actual barriers or with the barriers that are too thin for considering practical tunneling conduction. 
To examine the potential of $\mathrm{Mn_{3}Sn}$ as an electrode of the antiferromagnetic tunnel junctions,
it is required to evaluate the TMR effect using the $\mathrm{Mn_{3}Sn}$ electrode with thick enough barriers useful for applications.
Our present study, which ensures the tunneling transport,
will give a realistic reference point for further exploration of $\mathrm{Mn_{3}Sn}$-based MTJs.
\par
The remainder of this paper is as follows.
In Sec.~\ref{sec:method}, we discuss the construction of the MTJ with $\mathrm{Mn_{3}Sn}$ and $\mathrm{MgO}$,
and describe the theoretical methods to calculate the TMR effect in the MTJs and the material properties of $\mathrm{Mn_{3}Sn}$ and $\mathrm{MgO}$.
In Sec.~\ref{sec:results}, we present the results of first-principles calculations of the tunneling conductance in the MTJs as well as the spin splitting on the Fermi surface of $\mathrm{Mn_{3}Sn}$ and electron screening of $\mathrm{MgO}$.
Section~\ref{sec:summary} is devoted to the summary and outlook.
\section{System and Method}
\label{sec:method}
\subsection{Construction of magnetic tunnel junction with $\mathrm{Mn_{3}Sn}$ and $\mathrm{MgO}$}
\label{subsec:method_interface}
\begin{figure}[tbh]
	\centering
	\includegraphics[width=86mm]{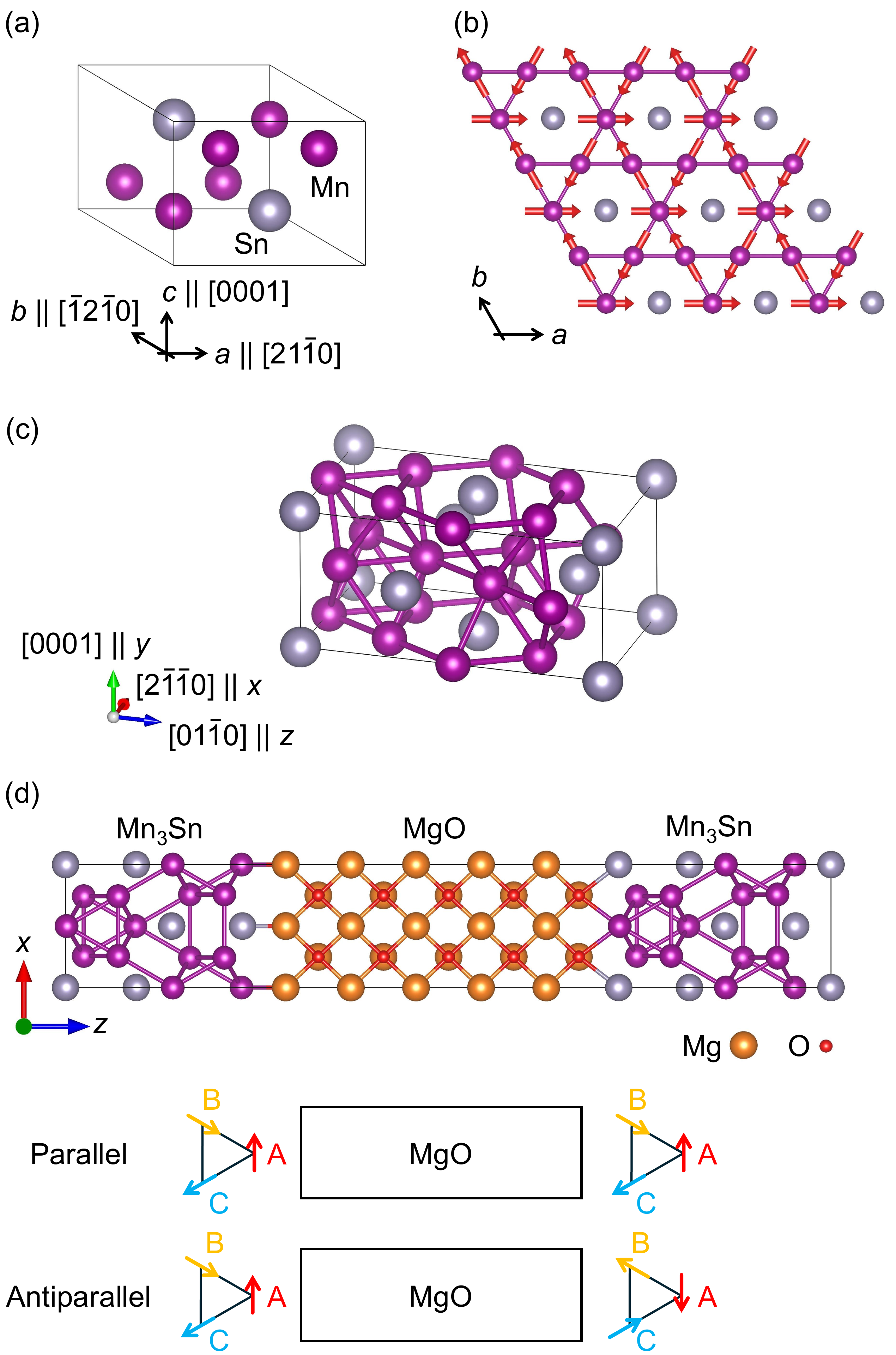}
	\caption{%
		(a) Crystal structure of $\mathrm{Mn_{3}Sn}$.
		(b) One of the two distinct kagome layers of $\mathrm{Mn_{3}Sn}$.
				Red arrows represent the magnetic moments carried by Mn ions in the inverse triangular lattice magnetic structure state.
		(c) Crystal structure of $\mathrm{Mn_{3}Sn}$ cut out as an orthorhombic shape, which is used for the calculations in this paper.
		(d) Crystal structure of the $\mathrm{Mn_{3}Sn}(01\bar{1}0)/\mathrm{MgO}(110)/\mathrm{Mn_{3}Sn}$ MTJ and illustration of the parallel and antiparallel configurations.
                The MTJ is viewed from $+y$-direction, which is parallel to the $\left[ 0001 \right]$-axis of $\mathrm{Mn_{3}Sn}$ (See also (c)).
				The electron tunnels through $z$-direction, parallel to the $\left[ 01\bar{1}0 \right]$-axis of $\mathrm{Mn_{3}Sn}$.
				Arrows in the schematic illustration denote the magnetic moments of $\mathrm{Mn}$-A/B/C ions.
	}
	\label{fig:MTJ}
\end{figure}
First, we discuss the construction of the MTJs.
We use $\mathrm{Mn_{3}Sn}$ with the $\mathrm{D0}_{19}$ structure as the electrode.
$\mathrm{D0}_{19}$-type $\mathrm{Mn_{3}Sn}$ has the hexagonal crystal structure whose space group is $P6_{3}/mmc$ (Fig.~\ref{fig:MTJ}(a)),
where Mn atoms form the bilayer kagome lattice.
The $a$-, $b$-, and $c$-axes are along $\left[ 2 1 \bar{1} 0 \right]$, $\left[ \bar{1} 2 \bar{1} 0 \right]$, and $\left[ 0 0 0 1 \right]$-axes, respectively.
We set the lattice constants $a = 5.665$~{\AA} and $c = 4.531$~{\AA} for $\mathrm{Mn_{3}Sn}$.
Mn atoms carry the magnetic moments of $\sim 3$~$\mu_{\mathrm{B}}$~\cite{Chen2024_PhysRevResearch_6_L032016}, 
and take the inverse triangular antiferromagnetic structure at room temperature as shown in Fig.~\ref{fig:MTJ}(b)~\cite{Nagamiya1982_SolidStateCommun_42_385,Tomiyoshi1982_JPhysSocJpn_51_803,Brown1990_JPhysCondensMatter_2_9409}.
This three-sublattice N\'{e}el ordered state breaks the macroscopic time-reversal symmetry,
which can be regarded as the ferroic order of the cluster octupole moment defined by the magnetic moments of the Mn ions~\cite{Suzuki2017_PhysRevB_95_094406,Suzuki2019_PhysRevB_99_174407}.
\par
We take $\mathrm{MgO}$ as the barrier material in the $\mathrm{Mn_{3}Sn}$-based MTJ,
which is one of the representative tunnel barriers suitable for MTJs~\cite{Parkin2004_NatMater_3_862,Yuasa2004_NatMater_3_868,Scheike2023_ApplPhysLett_122_112404},
and investigate the $\mathrm{Mn_{3}Sn}(01\bar{1}0)/\mathrm{MgO}(110)/\mathrm{Mn_{3}Sn}$ MTJ. 
For convenience in the following calculations, we cut $\mathrm{Mn_{3}Sn}$ as an orthorhombic cell as shown in Fig.~\ref{fig:MTJ}(c),
which contains twice the number of atoms as the original hexagonal unit cell.
Here, $x$-, $y$-, and $z$-axes of the orthorhombic cell are parallel to $\left[2\bar{1}\bar{1}0\right]$-, $\left[0001\right]$-, and $\left[01\bar{1}0\right]$-axes of $\mathrm{Mn_{3}Sn}$, respectively.
The crystal structure of the $\mathrm{Mn_{3}Sn}(01\bar{1}0)/\mathrm{MgO}(110)/\mathrm{Mn_{3}Sn}$ MTJ is shown in Fig.~\ref{fig:MTJ}(d).
To clarify how the intrinsic magnetic properties of bulk $\mathrm{Mn_{3}Sn}$ influence the TMR effect, we modeled an interface where $\sim 7.6$\% lattice mismatch is accommodated by applying an in-plane strain to the MgO lattice. 
Within this framework, we optimized the interfacial distance for several high-symmetry stacking configurations to identify the most stable structure.
\subsection{Calculation of tunneling conductance}
\label{subsec:method_tmr}
Next, we describe first-principles calculations of the TMR effect.
We calculate the TMR effect based on the scattering theory type approach using the Landauer--B\"{u}ttiker formula.
To perform the calculation, we divide the whole $\mathrm{Mn_{3}Sn}(01\bar{1}0)/\mathrm{MgO}(110)/\mathrm{Mn_{3}Sn}$ MTJ into three regions, 
two leads and the scattering region between the leads.
The lead region is $\mathrm{Mn_{3}Sn}$ shown in Fig.~\ref{fig:MTJ}(c).
The scattering region consists of the $\mathrm{MgO}(110)$ barrier and $\mathrm{Mn_{3}Sn}(01\bar{1}0)$ layers attached to its both sides as the buffer layers to smoothly connect the scattering region to the electrodes.
\par
We perform the density functional theory (DFT) calculation~\cite{Hohenberg1964_PhysRev_136_B864,Kohn1965_PhysRev_140_A1133} using the \textsc{Quantum ESPRESSO} (QE) package~\cite{Giannozzi2009_JPhysCondensMatter_21_395502,Giannozzi2017_JPhysCondensMatter_29_465901} to obtain the electronic structure for each of the aforementioned three regions.
The $\boldsymbol{k}$-mesh in the self-consistent field (scf) calculation is $6 \times 8 \times 4$ for the electrode and $6 \times 8 \times 1$ for the scattering region.
We use the ultrasoft pseudopotentials obtained from \textsc{pslibrary}~\cite{DalCorso2014_ComptMaterSci_95_337}.
The exchange correlation is taken into account by the Perdew--Burke--Ernzerhof generalized gradient approximation~\cite{Perdew1996_PhysRevLett_77_3865}.
The energy cutoff is 60~Ry for the wave-function and 480~Ry for the charge density.
We take the effects of the spin-orbit coupling into account unless otherwise noted.
When we perform the scf calculations of the scattering region, 
we impose the constraint of 1.0~Ry onto the initial magnetic structure with the magnetic moments of $3.0$~$\mu_{\mathrm{B}}$ for Mn-atoms and $0$~$\mu_{\mathrm{B}}$ for the other atoms to make the convergence faster.
We assume the inverse triangular magnetic structure shown in Fig.~\ref{fig:MTJ}(b) as the initial magnetic structure of $\mathrm{Mn_{3}Sn}$ and confirm that the inverse triangular lattice magnetic structure is realized overall after convergence,
although the leads and the scattering regions can have a tiny net magnetization.
\par
In discussing the magnetic configurations of MTJs, the parallel and antiparallel configurations are defined by the relative directions of the magnetic moments in the same sublattices between the left and right electrodes as schematically shown in Fig.~\ref{fig:MTJ}(d);
when we label three distinctive Mn ions as A, B, and C, 
each of the Mn-A, B, and C ions has a magnetic moment with parallel/antiparallel directions for the parallel/antiparallel configurations.
This corresponds to the relative directions of the cluster octupole moments between two magnetic electrodes. 
In the following, we take $x$ as the easy axis and thus the cluster octupole moment is aligned along $x$.
For the scf calculation of the antiparallel configuration, the scattering region is doubled, which is cut in half when we calculate the transmission in the following step.
\par
After the scf calculation, 
we perform the transmission calculation using the \textsc{pwcond} codes in the QE package~\cite{Smogunov2004_PhysRevB_70_045417,DalCorso2005_PhysRevB_71_115106,DalCorso2006_PhysRevB_74_045429}, 
which implements the calculation of the transmission and reflection probabilities based on the scattering theory~\cite{Choi1999_PhysRevB_59_2267}.
We calculate the total transmission, $T_{\text{tot}}$, by the Landauer--B\"{u}ttiker formula~\cite{Landauer1957_IBMJResDev_1_3,Landauer1970_PhilMag_21_863,Buttiker1986_PhysRevLett_57_1761,Buttiker1988_IBMJResDevelop_32_317} given as,
\begin{align}
	T_{\text{tot}} = \dfrac{1}{N_{\boldsymbol{k}_{\parallel}}} \sum_{\boldsymbol{k}_{\parallel}} T({\boldsymbol{k}_{\parallel}}).
	\label{eq:transmission_total}
\end{align}
Here, $T({\boldsymbol{k}_{\parallel}})$ is the partial transmission at in-plane $\boldsymbol{k}_{\parallel} = (k_{x}, k_{y})$ point perpendicular to the $z$-direction, which is along the conducting path.
The momentum dependence of this $T({\boldsymbol{k}_{\parallel}})$ reflects the spin polarization of tunneling electrons,
which is important to analyze the TMR effect, particularly the antiferromagnetic TMR effect.
The number of $\boldsymbol{k}_{\parallel}$-points is written as $N_{\boldsymbol{k}_{\parallel}}$.
We take $N_{\boldsymbol{k}_{\parallel}} = 101\times 101$ for plotting the momentum-resolved transmission,
and $N_{\boldsymbol{k}_{\parallel}} = 51\times 51$ for examining the MgO thickness dependence of the TMR effect.
The tunnel conductance, $G$, can be computed as
\begin{align}
	G = \dfrac{e^{2}}{h} T_{\text{tot}},
	\label{eq:conductance}
\end{align}
where $e$ is the elementary charge and $h$ is the Planck constant.
\par
To analyze the electronic states of the MTJ, 
we calculate the projected density of states of the scattering region with the parallel configuration after the scf calculation above and the subsequent non-scf (nscf) calculation. 
We take $15 \times 20 \times 1$ $\boldsymbol{k}$-points for the nscf calculation.
\subsection{Calculation of bulk properties of $\mathrm{Mn_{3}Sn}$ and $\mathrm{MgO}$}
\label{subsec:method_bulk}
In addition to the TMR calculation, we analyze the momentum dependence of the spin polarization on the Fermi surface of the electrode, $\mathrm{Mn_{3}Sn}$, and the complex band structure of the tunnel barrier, $\mathrm{MgO}$.
For $\mathrm{Mn_{3}Sn}$, we perform the nscf calculation with $6 \times 8 \times 4$ $\boldsymbol{k}$-points after the scf calculation of the $\mathrm{Mn_{3}Sn}$ lead with the settings mentioned in Sec~\ref{subsec:method_tmr}.
Then we perform the Wannierzation using the \textsc{wannier90} package~\cite{Pizzi2020_JPhysCondensMatter_32_165902}.
We use the $s$-, $p$-, and $d$-orbitals of $\mathrm{Mn}$ and the $s$- and $p$-orbitals of $\mathrm{Sn}$ for the Wannierization.
The spin polarization, $\boldsymbol{s}_{n}(\boldsymbol{k})$, for the eigenstate of $n$-th energy band with the momentum $\boldsymbol{k}$, $\ket{\psi_{n, \boldsymbol{k}}}$, given as,
\begin{align}
	\boldsymbol{s}_{n}(\boldsymbol{k})
= \braket{\psi_{n, \boldsymbol{k}}|\boldsymbol{s}|\psi_{n, \boldsymbol{k}}},
	\label{eq:wannier_spin}
\end{align}
is calculated from the Wannier-based tight-binding model with $201 \times 201\times 201$ $\boldsymbol{k}$-mesh.
Here, $\boldsymbol{s} = \frac{1}{2}\boldsymbol{\sigma}$ is the vector representation of the spin operators with the Pauli matrices, $\boldsymbol{\sigma}$.
\par
Additionally, in the ballistic transport of electrons in the TMR effect, 
the in-plane momentum $\boldsymbol{k}_{\parallel}$ dependence of the spin polarization of magnetic electrodes can affect the tunneling conductance.
To see the $\boldsymbol{k}_{\parallel}$-dependence of the spin polarization of $\mathrm{Mn_{3}Sn}$,
we also calculate the effective polarization projected onto the two-dimensional plane perpendicular to the conducting path, $\boldsymbol{p}(\boldsymbol{k}_{\parallel})$, which can be regarded as an extension of the in-plane polarization defined in the collinear magnets to noncollinear magnets~\cite{Gurung2024_NatCommun_15_10242}.
The expression of $\boldsymbol{p}(\boldsymbol{k}_{\parallel})$ is given as
\begin{align}
	\boldsymbol{p}(\boldsymbol{k}_{\parallel})
= \dfrac{\displaystyle\sum_{n} \int \boldsymbol{s}_{n}(\boldsymbol{k}) \ \delta(E_{n, \boldsymbol{k}} - E_{\mathrm{F}}) \ \mathrm{d}k_{z}}
	{\displaystyle\sum_{n} \int \left| \boldsymbol{s}_{n}(\boldsymbol{k}) \right| \ \delta(E_{n, \boldsymbol{k}} - E_{\mathrm{F}}) \ \mathrm{d}k_{z}},
	\label{eq:wannier_effective_polarization}
\end{align}
where $E_{n, \boldsymbol{k}}$ is the eigenenergy corresponding to $\ket{\psi_{n, \boldsymbol{k}}}$, and $E_{\text{F}}$ is the Fermi energy~\cite{note_deltafunc}.
\par
To analyze the tunneling decay of electrons through $\mathrm{MgO}(110)$ barrier in the MTJ,
we calculate the complex band structure, the energy eigenvalues with respect to the complex number of wave-vectors, 
of $\mathrm{MgO}$ with the \textsc{pwcond} codes.
At some specific energy points, we calculate the momentum dependence of the lowest decay rates inside $\mathrm{MgO}$ with $101 \times 101$ $\boldsymbol{k}_{\parallel}$-points.
The preceding spinless scf calculation is performed without including the spin-orbit coupling effect, 
where the $\boldsymbol{k}$-mesh is $6\times 8 \times 6$.
\section{Results and discussions}
\label{sec:results}
\subsection{Spin polarization on the Fermi surface of $\mathrm{Mn_{3}Sn}$}
\label{subsec:results_bulk}
\begin{figure}[tbh]
	\centering
	\includegraphics[width = 86mm]{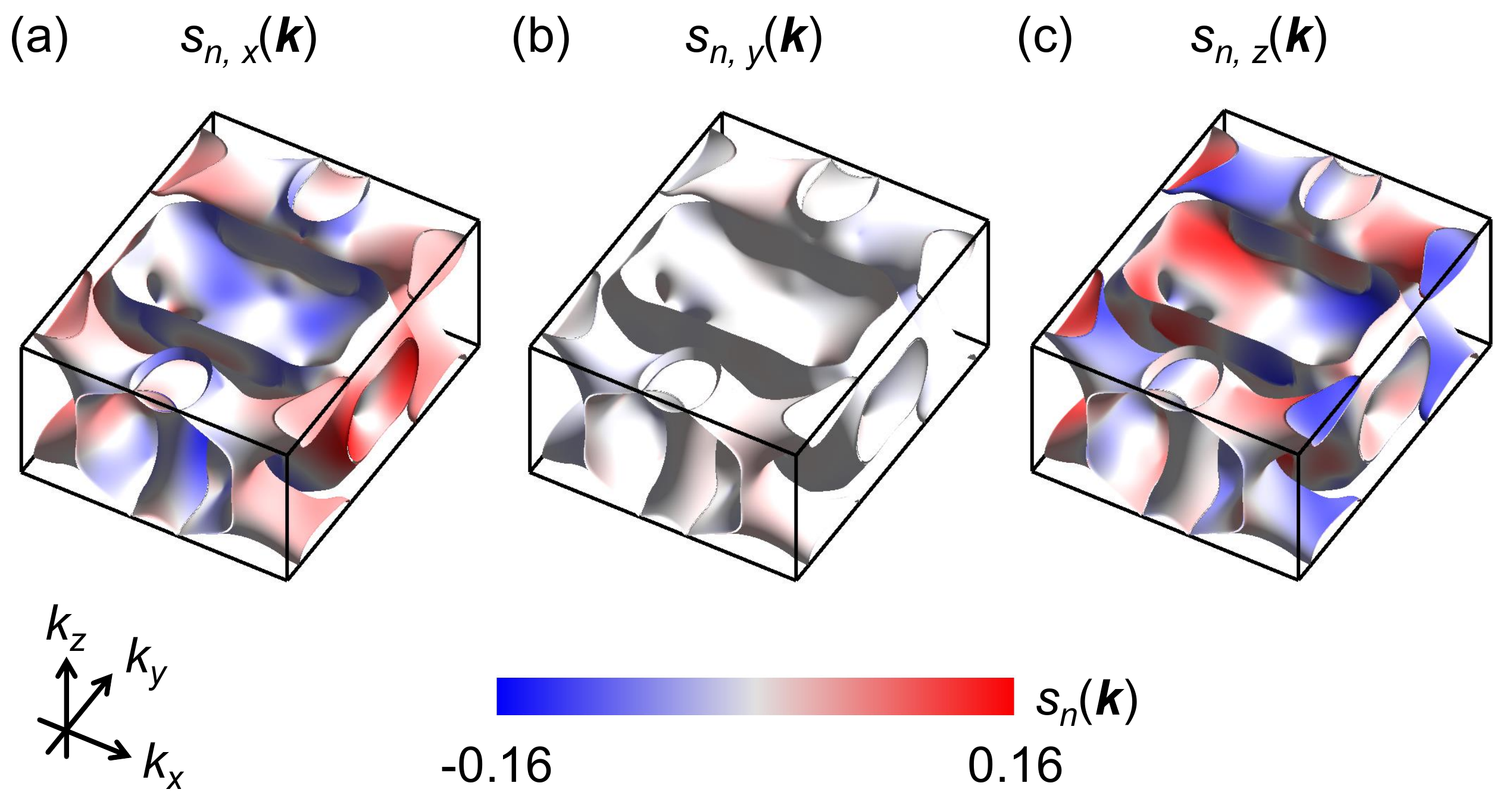}
	\caption{%
		Momentum-dependent spin polarization on the Fermi surfaces of $\mathrm{Mn_{3}Sn}$.
		(a) $s_{n, x}(\boldsymbol{k})$.
		(b) $s_{n, y}(\boldsymbol{k})$.
		(c) $s_{n, z}(\boldsymbol{k})$.
	}
	\label{fig:fermisurface}
\end{figure}
First, we discuss the momentum-dependent spin splitting in the bulk $\mathrm{Mn_{3}Sn}$ contributing to the TMR effect.
The Fermi surfaces of $\mathrm{Mn_{3}Sn}$ and the spin polarization are shown in Figs.~\ref{fig:fermisurface}(a)--\ref{fig:fermisurface}(c).
We find that the $s_{x}$- and $s_{z}$-components have a finite polarization,
while the $s_{y}$-component does not give a polarization distinctively.
\par
To investigate how the spin polarization in $\mathrm{Mn_{3}Sn}$ affects the TMR effect,
we discuss the $k_{z}$-dependence of $s_{n, x}(\boldsymbol{k})$ and $s_{n, z}(\boldsymbol{k})$.
This is because we should consider all the contributions of the several Bloch states with different momentum $k_{z}$ at each $\boldsymbol{k}_{\parallel}$-point when we discuss the tunneling transport along the $z$-direction.
We examine how the $s_{x}$ and $s_{z}$ components are transformed under the symmetry operations for the antiferromagnetic state of $\mathrm{Mn_{3}Sn}$~\cite{Suzuki2017_PhysRevB_95_094406,Dong2022_PhysRevLett_128_197201},
particularly focusing on the operations transforming $k_{z}$ to $-k_{z}$.
Specifically, here we pick up following three symmetry operations, $\left\{ PC_{2x} | \boldsymbol{0} \right\}$, $\left\{ TPC_{2z} | \boldsymbol{\tau} \right\}$, and $\left\{ P | \boldsymbol{0} \right\}$,
Here, $P$ is the spatial inversion operator, $C_{2\mu}$ ($\mu = x, y$) is the two-fold rotation operator along $\mu$-axis,
$T$ is the time-reversal operator,
and $\boldsymbol{0}$ and $\boldsymbol{\tau}$ represent the translational operator by $\left( x, y, z \right) = \left( 0, 0, 0 \right)$ and $\left( 0, c/2, 0 \right)$, respectively.
The transformation of the momentum and the spin under these three symmetry operations is summarized in Table~\ref{tbl:symmetry}.
Then, we obtain the following relations;
\begin{align}
	\begin{cases}
		s_{x}(k_{x}, k_{y}, k_{z}) & = s_{x}(k_{x}, k_{y}, -k_{z}) \\
		s_{z}(k_{x}, k_{y}, k_{z}) & = -s_{z}(k_{x}, k_{y}, -k_{z}) \\
	\end{cases}.
	\label{eq:spinpolarization_sym}
\end{align}
\begin{table}[tb]
	\centering
	\caption{%
		List of the symmetry operations for the antiferromagnetic state of $\mathrm{Mn_{3}Sn}$ and the transformation of the momentum and the spins under those operations.
		Note that we only list the symmetry operations essential for the present discussion on the spin-polarization,
		and there are other symmetry operations for $\mathrm{Mn_{3}Sn}$ (See Refs.~\cite{Suzuki2017_PhysRevB_95_094406,Dong2022_PhysRevLett_128_197201} for details).
	}
	\begin{tabular}{ccc} \hline
		Operation & Momentum: $(k_{x}, k_{y}, k_{z})$ & Spin: $(s_{x}, s_{y}, s_{z})$ \\ \hline
		$\left\{ PC_{2x} | \boldsymbol{0} \right\}$ & $(-k_{x}, k_{y}, k_{z})$ & $(s_{x}, -s_{y}, -s_{z})$ \\
		$\left\{ TPC_{2z} | \boldsymbol{\tau} \right\}$	& $(-k_{x}, -k_{y}, k_{z})$ & $(s_{x}, s_{y}, -s_{z})$ \\
		$\left\{ P | \boldsymbol{0} \right\}$ & $(-k_{x}, -k_{y}, -k_{z})$ & $(s_{x}, s_{y}, s_{z})$ \\ \hline
	\end{tabular}
	\label{tbl:symmetry}
\end{table}
These relations agree with the calculated momentum dependence of the spin polarization on the Fermi surfaces shown in Fig.~\ref{fig:fermisurface}.
From Eq.~(\ref{eq:spinpolarization_sym}), the contribution from $+k_{z}$ and $-k_{z}$ cancels out for $s_{z}$-component in the bulk form,
and thus $s_{z}$-component does not largely contribute to generating the difference in transmission between parallel and antiparallel configurations when we construct the MTJ.
By contrast, the $s_{x}$-component can take a finite value even when we take account of all of the momentum $k_{z}$ at each $\boldsymbol{k}_{\parallel}$-point since $s_{x}$ is symmetric with respect to the $k_{z} = 0$ plane.
Therefore, the TMR effect in the $\mathrm{Mn_{3}Sn}(01\bar{1}0)/\mathrm{MgO}(110)/\mathrm{Mn_{3}Sn}$ MTJ will occur dominantly due to the spin polarization of the $x$-component, 
i.e., the spin polarization along the easy axis of the cluster octupole moment.
\par
In addition, we also briefly discuss the $k_{x}$ and $k_{y}$-dependence of the $s_{x}$- and $s_{z}$-components;
again using the symmetry operations shown in Table~\ref{tbl:symmetry},
we obtain
\begin{align}
	\begin{cases}
		s_{x}(k_{x}, k_{y}, k_{z}) & = s_{x}(-k_{x}, k_{y}, k_{z}) \\
		s_{x}(k_{x}, k_{y}, k_{z}) & = s_{x}(k_{x}, -k_{y}, k_{z}) \\
		s_{z}(k_{x}, k_{y}, k_{z}) & = -s_{z}(-k_{x}, k_{y}, k_{z}) \\
		s_{z}(k_{x}, k_{y}, k_{z}) & = s_{z}(k_{x}, -k_{y}, k_{z}) \\
	\end{cases}.
	\label{eq:spinpolarization_sym_xy}
\end{align}
Namely, the $s_{x}$-component is symmetric with respect to the $k_{x} = 0$ and $k_{y} = 0$ planes,
and the $s_{z}$-component is antisymmetric with respect to the $k_{x} = 0$ plane and symmetric with respect to the $k_{y} = 0$ plane.
As well as the symmetry about $k_{z}$,
these relations on $k_{x}$ and $k_{y}$ are consistent with the numerical results shown in Fig.~\ref{fig:fermisurface}.
\par
\begin{figure}[tbh]
	\centering
	\includegraphics[width = 86mm]{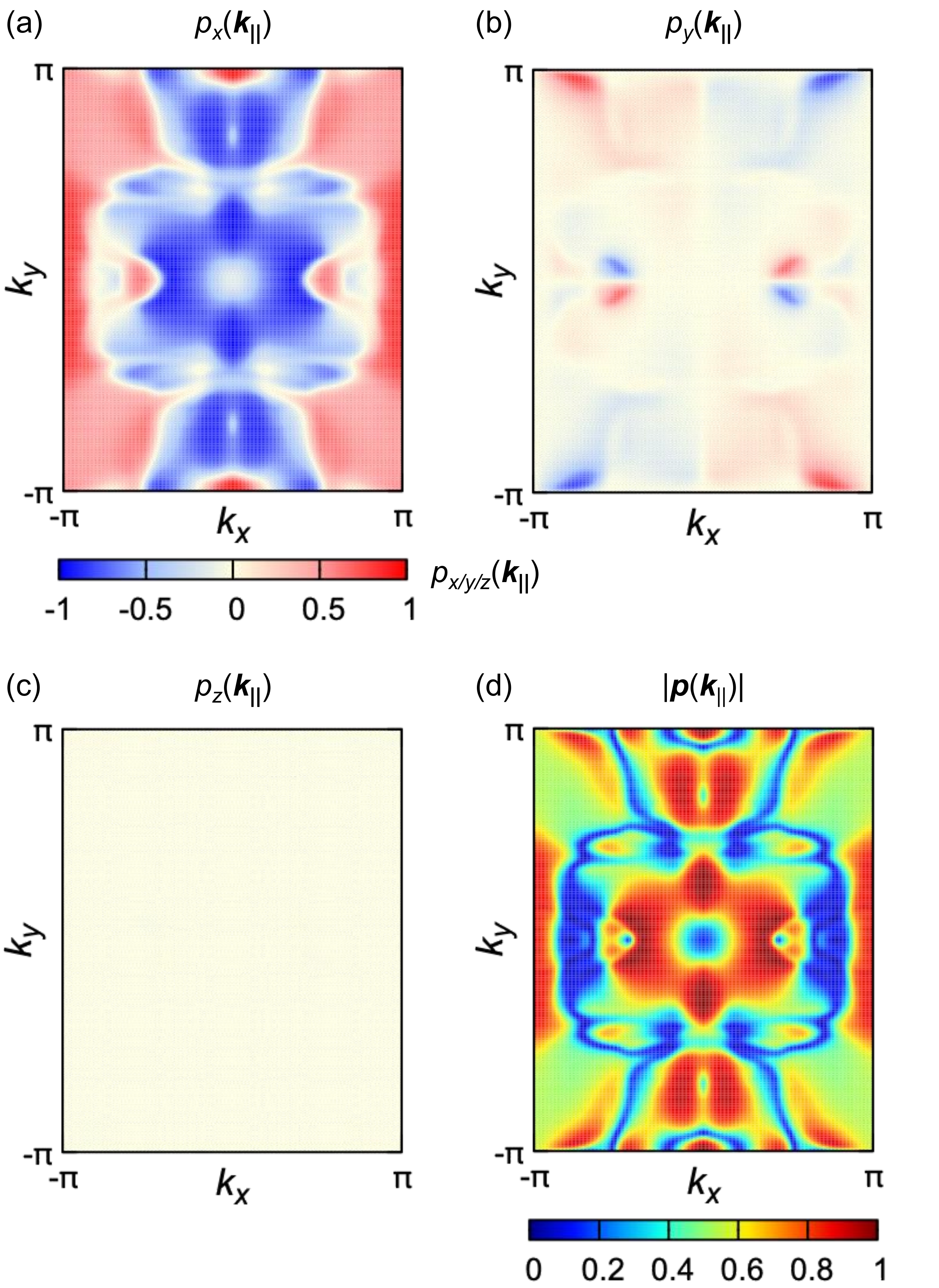}
	\caption{%
		Momentum dependence of the effective polarization of $\mathrm{Mn_{3}Sn}$ projected onto $xy$-plane, $\boldsymbol{p}(\boldsymbol{k}_{\parallel})$ (Eq.~(\ref{eq:wannier_effective_polarization})).
		(a) $p_{x}(\boldsymbol{k}_{\parallel})$.
		(b) $p_{y}(\boldsymbol{k}_{\parallel})$.
		(c) $p_{z}(\boldsymbol{k}_{\parallel})$.
		(d) The size of the effective polarization, $\left| \boldsymbol{p}(\boldsymbol{k}_{\parallel})\right|$.
	}
	\label{fig:2dpolarization}
\end{figure}
A more intuitive understanding of the spin polarization which can contribute to the TMR effect will be given by the effective polarization projected onto the two-dimensional plane $\boldsymbol{p}(\boldsymbol{k}_{\parallel})$ (Eq.~(\ref{eq:wannier_effective_polarization})),
where we sum up the contribution from the several conduction channels with different $k_{z}$ at each $\boldsymbol{k}_{\parallel}$-point.
Figures~\ref{fig:2dpolarization}(a)--\ref{fig:2dpolarization}(d) respectively show each component of $\boldsymbol{p}(\boldsymbol{k}_{\parallel})$, $p_{x}(\boldsymbol{k}_{\parallel})$, $p_{y}(\boldsymbol{k}_{\parallel})$, $p_{z}(\boldsymbol{k}_{\parallel})$,
and the size of the effective polarization, $\left| \boldsymbol{p}(\boldsymbol{k}_{\parallel})\right|$.
We find that $\left| \boldsymbol{p}(\boldsymbol{k}_{\parallel})\right|$ takes relatively large values at finite $\boldsymbol{k}_{\parallel}$ around $\boldsymbol{k}_{\parallel} = \boldsymbol{0}$,
which stems from the polarization of the $x$-component (Figs.~\ref{fig:2dpolarization}(a) and \ref{fig:2dpolarization}(d)).
This indicates that the $x$-component of the spin polarization can generate the TMR effect as we discussed above using the three-dimensional Fermi surfaces.
The distribution of $p_{x}(\boldsymbol{k}_{\parallel})$ is symmetric with respect to the $k_{x} = 0$ and $k_{y} = 0$ lines,
which reflects the symmetry of the spin polarization on the three-dimensional Fermi surfaces (See Fig.~\ref{fig:fermisurface}(a) and Eq.~(\ref{eq:spinpolarization_sym_xy})).
We also find that $p_{z}(\boldsymbol{k}_{\parallel})$ is small in the whole two-dimensional Brillouin zone as shown in Fig.~\ref{fig:2dpolarization}(c), due to the cancellation between the $+k_{z}$ component and the $-k_{z}$ component of $s_{n, z}(\boldsymbol{k})$ discussed above (Eq.~(\ref{eq:spinpolarization_sym})).
\subsection{Tunnel magnetoresistance effect}
\label{subsec:results_tmr}
\begin{figure*}[tbh]
	\centering
	\includegraphics[width=172mm]{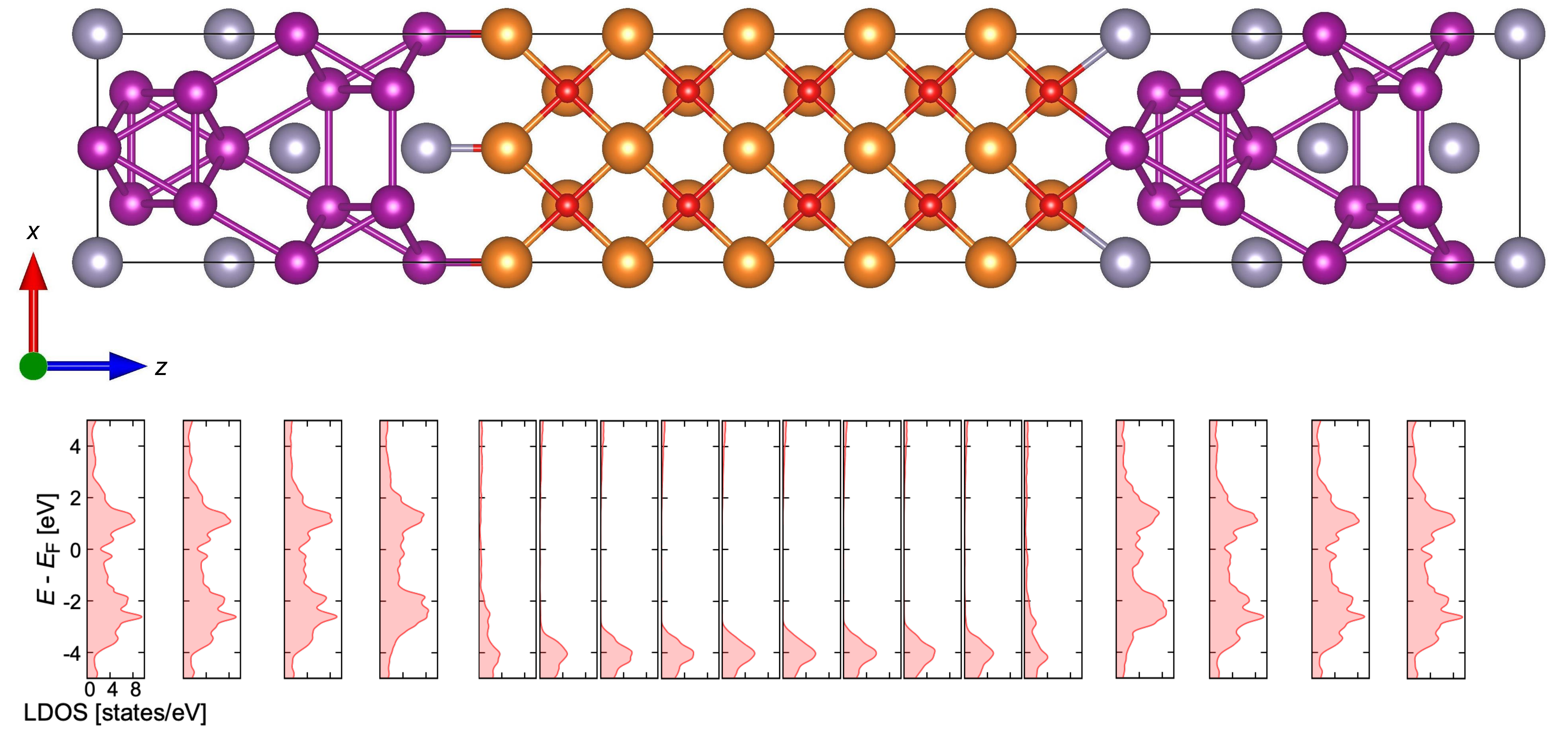}
	\caption{%
		Local density of states (LDOS) of each layer of the scattering region for the parallel configuration with 10 monolayers of $\mathrm{MgO}$.
		For the $\mathrm{Mn_{3}Sn}$ regions, the LDOS summed for several different layers with close $z$-coordinates is shown.
	}
	\label{fig:LDOS_layer}
\end{figure*}
\begin{figure}[tbh]
	\centering
	\includegraphics[width=86mm]{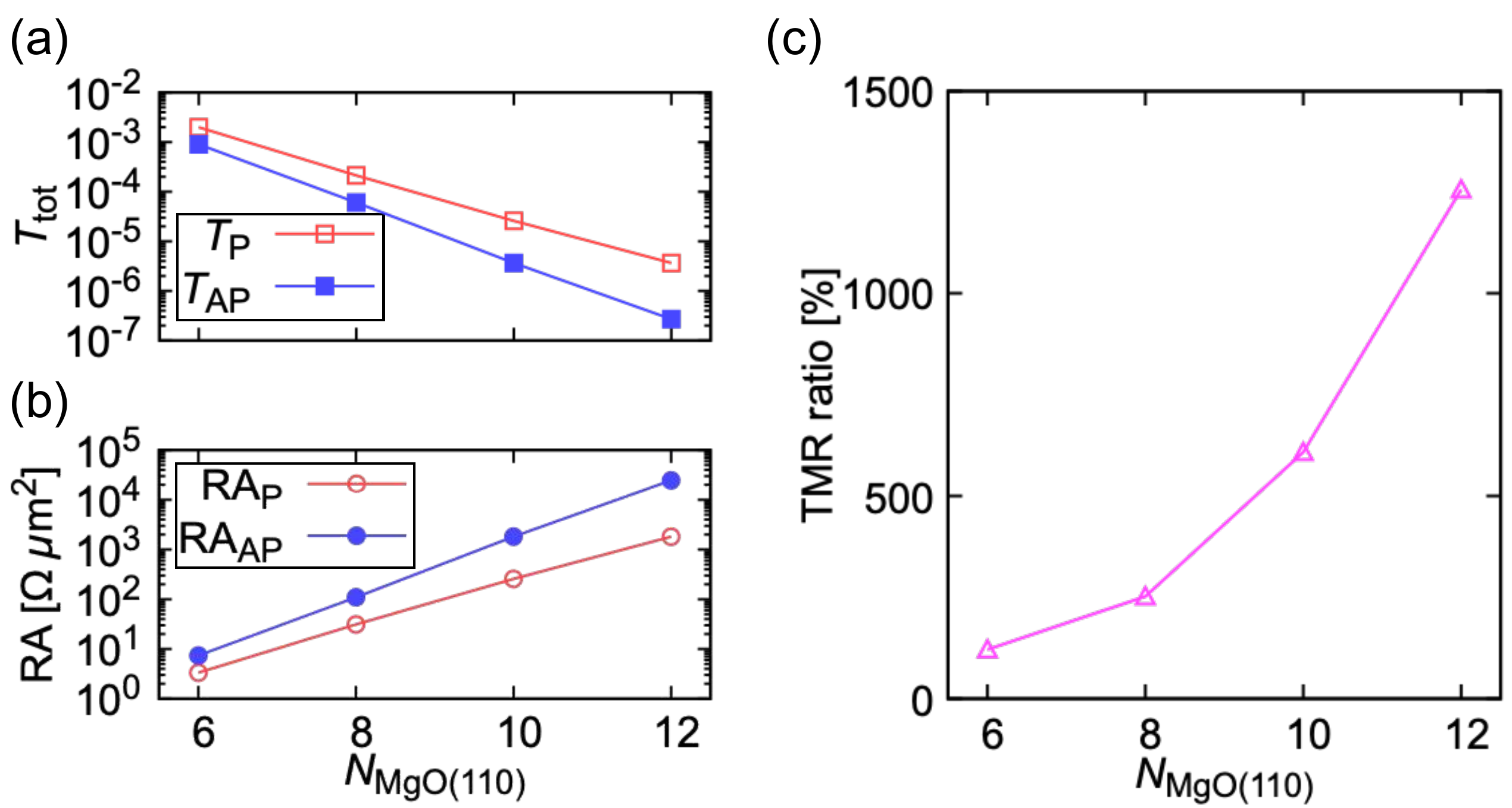}
	\caption{%
		(a) Total transmission for the parallel and antiparallel configurations in $\mathrm{Mn_{3}Sn}/\mathrm{MgO}/\mathrm{Mn_{3}Sn}$ magnetic tunnel junctions (MTJs) with respect to the number of $\mathrm{MgO}$ layers, $N_{\mathrm{MgO}(110)}$.
		(b) RA value.
		(c) MgO thickness dependence of the TMR ratio corresponding to (a).
	}
	\label{fig:TMR}
\end{figure}
\begin{figure*}[tbh]
	\centering
	\includegraphics[width=129mm]{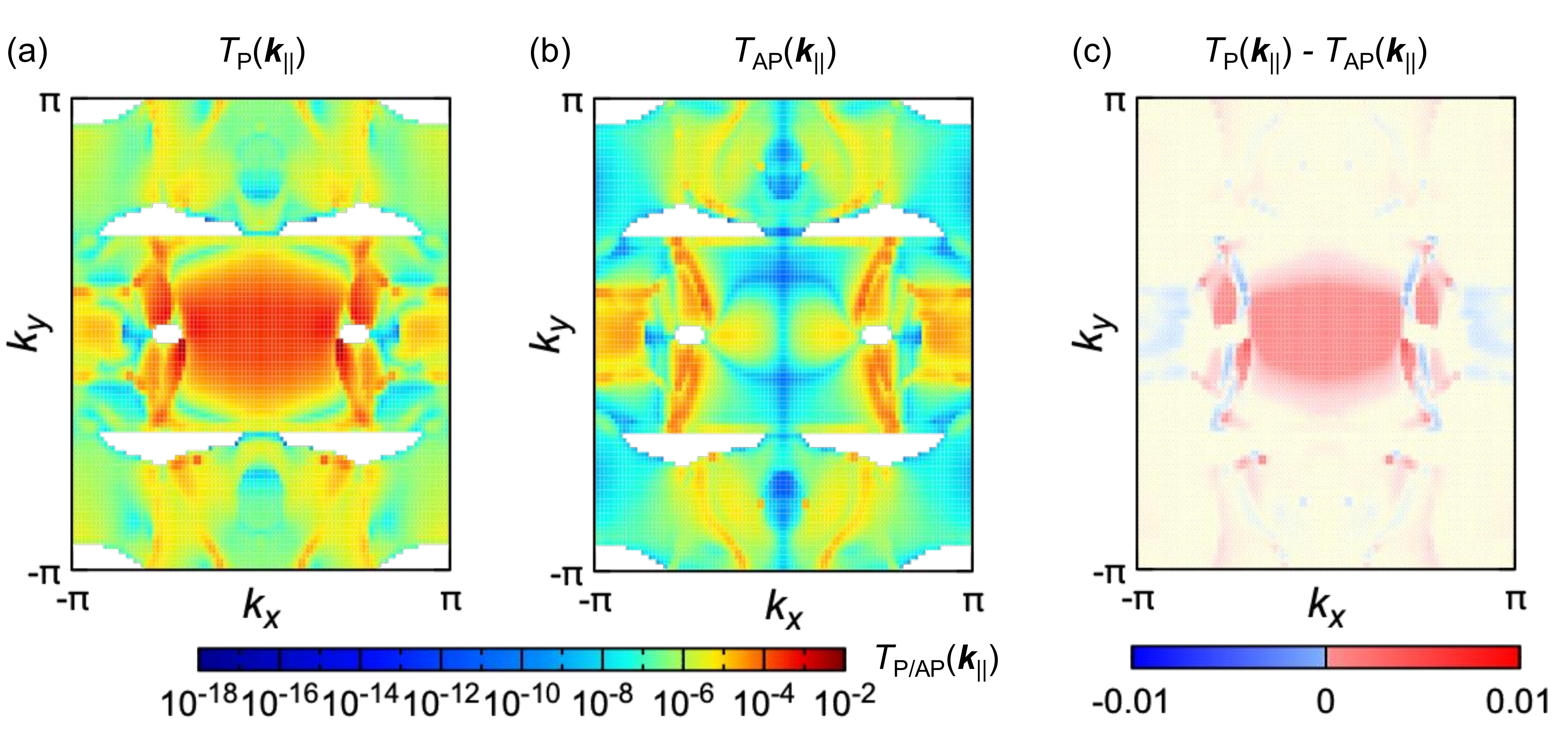}
	\caption{%
		(a), (b) Momentum-resolved transmission of the (a) parallel and (b) antiparallel configurations, $T_{\text{P}}(\boldsymbol{k}_{\parallel})$ and $T_{\text{AP}}(\boldsymbol{k}_{\parallel})$, 
		in the $\mathrm{Mn_{3}Sn}/\mathrm{MgO}/\mathrm{Mn_{3}Sn}$ MTJ with $N_{\mathrm{MgO}(110)} = 10$.
		(c) Difference in the transmission between parallel and antiparallel configurations, $T_{\text{P}}(\boldsymbol{k}_{\parallel})-T_{\text{AP}}(\boldsymbol{k}_{\parallel})$.
	}
	\label{fig:TMR_kres}
\end{figure*}
\begin{figure}[tbh]
	\centering
	\includegraphics[width=86mm]{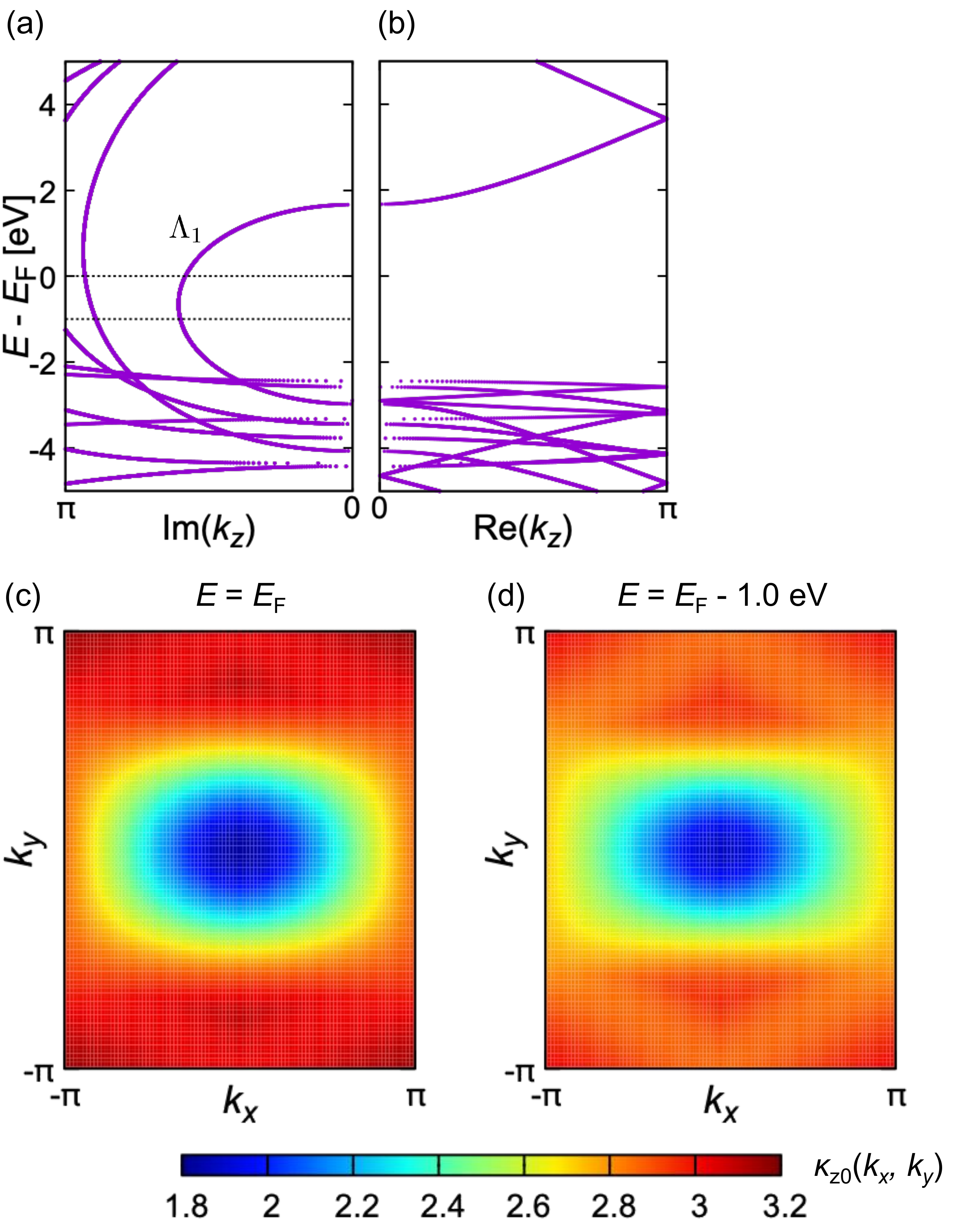}
	\caption{%
		(a), (b) Complex energy bands of $\mathrm{MgO}$ with $k_{x} = k_{y} = 0$ for (a) the pure imaginary wave vectors $\mathrm{Im}(k_{z})$ and (b) the real wave vectors $\mathrm{Re}(k_{z})$.
		Broken lines in (a) are the energies where the momentum dependence of the decay rate is shown in (c) and (d).
		The energy band with the lowest decay rate around $E_{\mathrm{F}}$ belongs to the $\Lambda_{1}$ irreducible representation.
		(c), (d) In-plane momentum dependence of the lowest decay rate of $\mathrm{MgO}$, $\kappa_{z0}(k_{x}, k_{y})$, 
        at (c) $E = E_{\mathrm{F}}$ and (d) $E = E_{\mathrm{F}} -1.0$~eV.
	}
	\label{fig:MgO}
\end{figure}
Next, we examine the $\mathrm{Mn_{3}Sn}(01\bar{1}0)/\mathrm{MgO}(110)/\mathrm{Mn_{3}Sn}$ MTJs.
Figure~\ref{fig:LDOS_layer} shows the local density of states (LDOS) inside and near the barrier layers of the scattering region with $N_{\mathrm{MgO}(110)} = 10$ for the parallel configuration.
The LDOS in the MgO barrier at the Fermi level is small enough, 
which ensures the electron transport discussed here is the tunneling one.
\par
Then we discuss the TMR effect in the $\mathrm{Mn_{3}Sn}(01\bar{1}0)/\mathrm{MgO}(110)/\mathrm{Mn_{3}Sn}$ MTJs.
We show the MgO thickness dependence of the total transmission for the parallel configuration, $T_{\text{P}}$, 
and that for the parallel configuration, $T_{\text{AP}}$, in Fig.~\ref{fig:TMR}(a).
We find that both of $T_{\text{P}}$ and $T_{\text{AP}}$ decrease exponentially as the number of MgO barrier layers increases,
which indicates that the transport is the tunneling one as well as the LDOS mentioned above.
We also show the $N_{\mathrm{MgO}(110)}$-dependence of the resistance-area product (RA) in Fig.~\ref{fig:TMR}(b),
calculated by $\text{RA} = A/G$,
where $A$ is the cross-section of the MTJ used in the calculation.
The RA value is $\lesssim$ 1--10 $\mathrm{k\Omega}\cdot\mathrm{\mu m^{2}}$ at $N_{\mathrm{MgO}(110)} = 12$,
which will be a reasonable value for the actual MTJ devices.
The corresponding TMR ratio, given by
$\text{[TMR ratio]} \ [\%] 
= \left( T_{\text{P}} - T_{\text{AP}} \right) / T_{\text{AP}}\times 100$,
is shown in Fig.~\ref{fig:TMR}(c),
which reaches $\gtrsim 1000$\% at $N_{\mathrm{MgO}(110)} = 12$.
\par
In Figs.~\ref{fig:TMR_kres}(a) and \ref{fig:TMR_kres}(b), 
we show the momentum-dependence of the transmission in the $\mathrm{Mn_{3}Sn}/\mathrm{MgO}/\mathrm{Mn_{3}Sn}$ MTJ with $N_{\mathrm{MgO}(110)} = 10$ for the parallel and antiparallel configurations, $T_{\text{P}}(\boldsymbol{k}_{\parallel})$ and $T_{\text{AP}}(\boldsymbol{k}_{\parallel})$, respectively.
We find a large difference in the transmission around $\boldsymbol{k}_{\parallel} = \boldsymbol{0}$,
which dominantly contributes to the TMR effect in the $\mathrm{Mn_{3}Sn}(01\bar{1}0)/\mathrm{MgO}(110)/\mathrm{Mn_{3}Sn}$ MTJs.
We also show the difference of the partial transmissions between the parallel and antiparallel configurations at each $\boldsymbol{k}_{\parallel}$-point, $T_{\text{P}}(\boldsymbol{k}_{\parallel}) - T_{\text{AP}}(\boldsymbol{k}_{\parallel})$, in Fig.~\ref{fig:TMR_kres}(c).
By focusing on the $\boldsymbol{k}_{\parallel}$-dependence of the difference $T_{\text{P}}(\boldsymbol{k}_{\parallel}) - T_{\text{AP}}(\boldsymbol{k}_{\parallel})$, 
we can clearly see which $\boldsymbol{k}_{\parallel}$-point largely contributes to the TMR effect, 
since the TMR ratio can be written as,
\begin{align}
	\text{[TMR ratio]} \ [\%] 
	& = \dfrac{T_{\text{P}} - T_{\text{AP}}}{T_{\text{AP}}} \times 100 \notag \\
	& = \sum_{\boldsymbol{k}_{\parallel}}\dfrac{T_{\text{P}}(\boldsymbol{k}_{\parallel}) - T_{\text{AP}}(\boldsymbol{k}_{\parallel})}{T_{\text{AP}}} \times 100.
	\label{eq:tmr_ratio}
\end{align}
We can observe that $T_{\text{P}}(\boldsymbol{k}_{\parallel}) - T_{\text{AP}}(\boldsymbol{k}_{\parallel})$ becomes larger around $\boldsymbol{k}_{\parallel} \sim \boldsymbol{0}$,
which contributes to the TMR effect in the $\mathrm{Mn_{3}Sn}/\mathrm{MgO}/\mathrm{Mn_{3}Sn}$ MTJs,
in a more pronounced form.
This difference in the transmission is caused by the momentum-dependent spin polarization of $\mathrm{Mn_{3}Sn}$ as shown in Figs.~\ref{fig:fermisurface} and~\ref{fig:2dpolarization};
the distribution in Fig.~\ref{fig:TMR_kres}(c) roughly reflects the effective polarization, namely, 
$\left| \boldsymbol{p}(\boldsymbol{k}_{\parallel})\right|$ takes relatively large values at finite $\boldsymbol{k}_{\parallel}$ around $\boldsymbol{k}_{\parallel} = \boldsymbol{0}$.
\par
In addition to the spin polarization of $\mathrm{Mn_{3}Sn}$ in the momentum space, this momentum dependence of the transmission is also supported by the filtering of the tunneling electrons in the $\mathrm{MgO}$ barrier.
We can clearly see the contribution of $\mathrm{MgO}$ by comparing the transmission in the $\mathrm{Mn_{3}Sn}(01\bar{1}0)/\mathrm{MgO}(110)/\mathrm{Mn_{3}Sn}$ MTJ and that in the $\mathrm{Mn_{3}Sn}(01\bar{1}0)/\text{vacuum}/\mathrm{Mn_{3}Sn}$ MTJ (See Appendix~\ref{sec:appendix_vacuum} for the details of the $\mathrm{Mn_{3}Sn}(01\bar{1}0)/\text{vacuum}/\mathrm{Mn_{3}Sn}$ MTJ).
In the $\mathrm{Mn_{3}Sn}(01\bar{1}0)/\text{vacuum}/\mathrm{Mn_{3}Sn}$ MTJ,
the momentum dependence of the transmission totally changes by removing the $\mathrm{MgO}$ layers,
which is due to the absence of the screening effect of the $\mathrm{MgO}$ barrier in the following discussion.
We can understand the screening effect of $\mathrm{MgO}$ by its complex band structure.
In Fig.~\ref{fig:MgO}(a), we show the imaginary part of the energy bands at $\boldsymbol{k}_{\parallel} = \boldsymbol{0}$.
When we extend the momentum of the Bloch wave functions to the complex number, $\bar{k}_{z}$, 
as $\bar{k}_{z} = k_{z} + \mathrm{i}\kappa_{z}$ with the real numbers $k_{z}$ and $\kappa_{z}$,
the imaginary part can express the decay of the electron wave function inside $\mathrm{MgO}$.
This can be more explicitly understood by rewriting the Bloch wave function as $\mathrm{e}^{\mathrm{i}\bar{k}_{z}z} = \mathrm{e}^{\mathrm{i}k_{z} z}\mathrm{e}^{- \kappa_{z} z}$.
Then, the imaginary part, $\kappa_{z}$, describes the exponential decay of the tunneling electrons.
In general, there are multiple $\kappa_{z}$ for each energy point as shown in Fig.~\ref{fig:MgO}(a), 
but we only have to focus on the imaginary band with the smallest $\kappa_{z}$, $\kappa_{z0}$, which should dominantly contribute to the tunneling transport at each $\boldsymbol{k}_{\parallel}$-point and energy,
considering that the tunneling electrons with larger $\kappa_{z}$ decay exponentially faster than the electrons with $\kappa_{z0}$.
We show the lowest decay rate $\kappa_{z0}$ at each $\boldsymbol{k}_{\parallel}$-point at $E = E_{\mathrm{F}}$ and $E_{\mathrm{F}} -1.0$~eV in Figs.~\ref{fig:MgO}(c) and \ref{fig:MgO}(d), respectively.
Here the energy point $E_{\mathrm{F}} -1.0$~eV is chosen by comparing the LDOS of $\mathrm{MgO}$ inside the scattering region (see Fig.~\ref{fig:LDOS_layer}) and the DOS of bulk $\mathrm{MgO}(110)$; 
the energy shifts by about $-1$~eV.
Both in Figs.~\ref{fig:MgO}(c) and \ref{fig:MgO}(d),
we clearly see the lowest decay rate takes a smaller value around $\boldsymbol{k}_{\parallel} \sim \boldsymbol{0}$.
This momentum dependence is consistent with the transmission properties through MTJ as shown in Fig.~\ref{fig:TMR_kres}(c).
\par
We note the symmetry of the electrons which mainly generate the TMR effect.
In particular, we focus on the $\boldsymbol{k}_{\parallel} = \boldsymbol{0}$ point since the $\boldsymbol{k}_{\parallel} = \boldsymbol{0}$ point has a dominant contribution to the TMR effect as we have discussed it above.
Among the energy bands with $k_{x} = k_{y} = 0$ (Figs.~\ref{fig:MgO}(a) and \ref{fig:MgO}(b)),
the energy band with the lowest decay rate around $E_{\mathrm{F}}$,
which is central to the TMR effect,
has the $\Lambda_{1}$ symmetry.
This $\Lambda_{1}$ symmetry state shares the common symmetry with the $\Delta_{1}$ symmetry state as discussed in the Fe(001)/MgO(001)/Fe MTJs~\cite{Butler2001_PhysRevB_63_054416}, 
in that both belong to the totally symmetric irreducible representations.
Namely, in the $\mathrm{Mn_{3}Sn}(01\bar{1}0)/\mathrm{MgO}(110)/\mathrm{Mn_{3}Sn}$ MTJs, 
the electrons with the totally symmetric irreducible representation will play a main role in the TMR effect as in the Fe(001)/MgO(001)/Fe MTJs,
though we may need further discussion on the detailed contribution of the totally symmetric states in $\mathrm{Mn_{3}Sn}$.
\par
So far we have examined two magnetic configurations of the MTJ, the parallel and antiparallel configurations.
More generally, one may consider the angular dependence of the magnetic moments in addition to those two alignments.
Let us consider the case where we measure the relative angle of the magnetic moments on the same sublattices in two distinct magnetic layers, $\theta$, with reference to the parallel configuration.
Namely, we regard the parallel and antiparallel configurations as the $\theta = 0^{\circ}$ and $180^{\circ}$ states, respectively.
Then, from the parallel to the antiparallel configurations, the tunneling conductance will decrease as a cosine function of $\theta$.
This cosine-type behavior has been discussed in the $\mathrm{Mn_{3}Sn}(0001)/\text{vacuum}/\mathrm{Mn_{3}Sn}(0001)$ antiferromagnetic MTJ~\cite{Dong2022_PhysRevLett_128_197201} as well as the conventional ferromagnetic MTJs~\cite{Slonczewski1989_PhysRevB_39_6995}.
\section{Summary}
\label{sec:summary}
In summary, we have studied first-principles calculations of the tunnel magnetoresistance (TMR) effect in the $\mathrm{Mn_{3}Sn}(01\bar{1}0)/\mathrm{MgO}(110)/\mathrm{Mn_{3}Sn}$ magnetic tunnel junction (MTJ).
This is an optimal structure for the switching of the magnetic configurations with the spin-orbit torque triggered by the electric current.
We have ensured the tunneling transport in the MTJs using a reasonably thick barrier and found that the calculated TMR ratio reaches more than 1000\%.
Examining the momentum-dependent transmission through MTJ combined with the spin splitting of bulk $\mathrm{Mn_{3}Sn}$ and the decaying properties of tunneling electrons inside $\mathrm{MgO}$, 
we have confirmed that the TMR effect occurs due to the spin splitting in the momentum space of $\mathrm{Mn_{3}Sn}$ and the electron filtering by the $\mathrm{MgO}$ barrier.
\par
For $\mathrm{Mn_{3}Sn}$, the spintronic functions have been intensively investigated toward applications, such as electrical switching, domain wall propagation, or exchange bias effect~\cite{Tsai2020_Nature_580_608,Pal2022_SciAdv_8_eabo5930,Higo2022_Nature_607_474,Yoon2023_NatMater_22_1106,Asakura2024_AdvMater_36_2400301,Wu2024_NatCommun_15_4305,Zhang2025_AdvSci_12_2417621,Asakura2025_NanoLett_25_10294,Takeuchi2025_Science_389_8301}.
Our study suggests that $\mathrm{Mn_{3}Sn}$ can also serve as a promising core of the MTJ devices for electric readout.
\begin{acknowledgments}
KT thanks Shinji Miwa and Koji Inukai for communications.
This work was supported by JST-Mirai Program (Grant No.~JPMJMI20A1), 
JST-CREST (Grant No.~JPMJCR23O4), 
JST-ASPIRE (Grant No.~JPMJAP2317), 
JSPS-KAKENHI (Grant No.~JP21H04437, No.~JP21H04990, No.~JP22H00290, No.~JP23H04869, No.~JP24K00581, No.~JP25K17935, No.~JP25K21684, No.~JP25H00420, No.~JP25H01252), 
and the RIKEN TRIP initiative (RIKEN Quantum, AGIS, Many-body Electron Systems).
This work was supported by JSR Corporation via JSR-UTokyo Collaboration Hub, CURIE. 
Part of the numerical calculations in this study was performed by using the Supercomputer in the Institute for Solid State Physics, University of Tokyo.
We use the \textsc{VESTA} software~\cite{Momma2011_JApplCryst_44_1272} to visualize the crystal structures with the aid of the \textsc{XCrySDen} software~\cite{Kokalj1999_JMolGraphicsModelling_17_176} and the \textsc{FermiSurfer} software to visualize the Fermi surfaces~\cite{Kawamura2019_ComputPhysCommun_239_197}.
\end{acknowledgments}
\appendix
\section{Calculation of the tunnel magnetoresistance effect in the $\mathrm{Mn_{3}Sn}(01\bar{1}0)/\text{vacuum}/\mathrm{Mn_{3}Sn}$ magnetic tunnel junction}
\label{sec:appendix_vacuum}
\begin{figure*}[tbh]
	\centering
	\includegraphics[width=129mm]{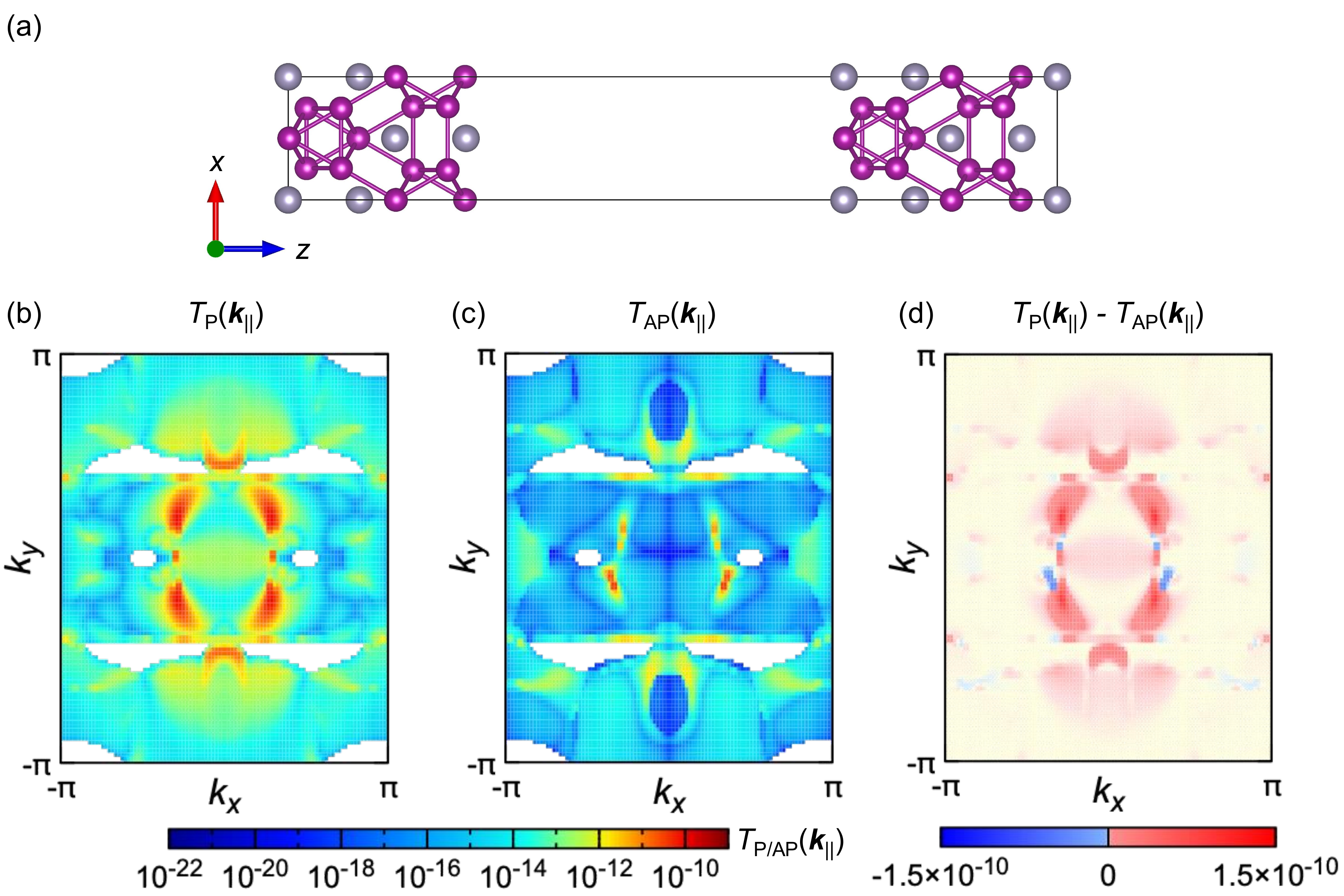}
	\caption{%
		(a) Crystal structure of the $\mathrm{Mn_{3}Sn}(01\bar{1}0)/\text{vacuum}/\mathrm{Mn_{3}Sn}$ magnetic tunnel junction (MTJ).
		(b), (c) Momentum-resolved transmission for the (b) parallel and (c) antiparallel configurations, 
		$T_{\text{P}}(\boldsymbol{k}_{\parallel})$ and $T_{\text{AP}}(\boldsymbol{k}_{\parallel})$, 
		in the $\mathrm{Mn_{3}Sn}(01\bar{1}0)/\text{vacuum}/\mathrm{Mn_{3}Sn}$ MTJ.
		The distance between two $\mathrm{Mn_{3}Sn}$ is the same with that in $\mathrm{Mn_{3}Sn}(01\bar{1}0)/\mathrm{MgO}/\mathrm{Mn_{3}Sn}$ MTJ with 10 monolayers of MgO.
		(d) Difference in the transmission between parallel and antiparallel configurations, $T_{\text{P}}(\boldsymbol{k}_{\parallel})-T_{\text{AP}}(\boldsymbol{k}_{\parallel})$.
	}
	\label{fig:TMR_vacuum}
\end{figure*}
In this Appendix, we show the momentum dependence of the transmission in the $\mathrm{Mn_{3}Sn}/\text{vacuum}/\mathrm{Mn_{3}Sn}$ MTJ to confirm the tunneling properties with $\mathrm{Mn_{3}Sn}$ electrodes.
The crystal structure of the $\mathrm{Mn_{3}Sn}/\text{vacuum}/\mathrm{Mn_{3}Sn}$ MTJ is shown in Fig.~\ref{fig:TMR_vacuum}(a).
Here, the distance between two $\mathrm{Mn_{3}Sn}$ electrodes is matched to that in $\mathrm{Mn_{3}Sn}/\mathrm{MgO}/\mathrm{Mn_{3}Sn}$ MTJ with 10 monolayers of MgO.
Figures~\ref{fig:TMR_vacuum}(b) and \ref{fig:TMR_vacuum}(c) show the momentum dependence of the transmission $T(\boldsymbol{k}_{\parallel})$ through $\mathrm{Mn_{3}Sn}/\text{vacuum}/\mathrm{Mn_{3}Sn}$ MTJ for the parallel and antiparallel configurations, 
$T_{\text{P}}(\boldsymbol{k}_{\parallel})$ and $T_{\text{AP}}(\boldsymbol{k}_{\parallel})$, respectively. 
Here we take $N_{\boldsymbol{k}_{\parallel}} = 101\times 101$.
We can see a large transmission at the $\boldsymbol{k}_{\parallel} \neq \boldsymbol{0}$ region.
We also show the difference of the partial transmissions between the parallel and antiparallel configurations $T_{\text{P}}(\boldsymbol{k}_{\parallel}) - T_{\text{AP}}(\boldsymbol{k}_{\parallel})$ in Fig.~\ref{fig:TMR_vacuum}(d).
We observe that the difference becomes large at the $\boldsymbol{k}_{\parallel} \neq \boldsymbol{0}$ region.
This result reflects the momentum dependent spin splitting of $\mathrm{Mn_{3}Sn}$, particularly the $x$-component of spin polarization (Figs.~\ref{fig:fermisurface}(a) and \ref{fig:2dpolarization}(a)), as discussed in Sec.~\ref{subsec:results_bulk}.
%


\begin{thebibliography}{104}
\bibitem [{not()}]{note_deltafunc}%
  \BibitemOpen
  \href@noop {} {}\bibinfo {note} {In the numerical integration, we approximate
  the delta function by the Gaussian distribution with a small finite smearing
  and consider only the energy bands crossing $E_{\mathrm{F}}$ for
  integrand.}\BibitemShut {Stop}%
\end{thebibliography}
\end{document}